\documentclass[lettersize,journal]{IEEEtran}
\usepackage{amsmath,amsfonts}
\usepackage{amssymb}
\usepackage{algorithmic}
\usepackage{algorithm}
\usepackage{array}
\usepackage[caption=false,font=footnotesize]{subfig}
\usepackage{textcomp}
\usepackage{stfloats}
\usepackage{url}
\usepackage{verbatim}
\usepackage{graphicx}
\usepackage{gensymb}
\usepackage{cite}
\usepackage{lzqsymbol}
\usepackage{amsthm}
\usepackage{booktabs} 
\usepackage{multirow}

\usepackage{xcolor}

\usepackage{tcolorbox}

\begin{document}

\title{Parallax QAMA: Novel Downlink Multiple Access for MISO Systems with Simple Receivers \\
\thanks{The work of J. Huang, M. Zhao and J. Zhu was supported by the National Key R\&D Program of China under
Grant 2022YFB2902005.}}

\author{
Jie~Huang,
Ming~Zhao,~\IEEEmembership{Member,~IEEE},
~Shengli~Zhou,~\IEEEmembership{Fellow,~IEEE}, 
Ling~Qiu,~\IEEEmembership{Member,~IEEE},
Jinkang~Zhu,~\IEEEmembership{Life~Member,~IEEE,}
\thanks{ Jie Huang, Ming Zhao, and Ling Qiu are with 
CAS Key Laboratory of Wireless-Optical Communications,
University of Science and Technology of China (USTC),
Hefei, Anhui Province, 230027, P. R. China (e-mail: ziven@mail.ustc.edu.cn, zhaoming@ustc.edu.cn, lqiu@ustc.edu.cn).} 
\thanks{Shengli Zhou is with Department of Electrical and Computer
Engineering, University of Connecticut, 371 Fairfield Way, Unit 4157, Storrs, CT
06269-4157 (e-mail: shengli.zhou@uconn.edu).}
\thanks{ Jinkang Zhu is with Dept. of Electronic Engineering and
Information Science, University of Science and Technology of China (USTC),
Hefei, Anhui Province, 230027, P. R. China (e-mail: jkzhu@ustc.edu.cn).
}
}

\maketitle
\setcounter{footnote}{0}
\begin{abstract}
In this paper, we propose a novel downlink multiple access system with a multi-antenna transmitter and two single-antenna receivers, inspired by the underlying principles of hierarchical quadrature amplitude modulation (H-QAM) based multiple access (QAMA) and space-division multiple access (SDMA).
In the proposed scheme, coded bits from two users are split and assigned to one shared symbol and two private symbols carried by different beams.
Based on joint symbol mapping of H-QAM constellations and phase-aligned precoding at the transmitter, each receiver observes a different H-QAM constellation with Gray mapping, a unique parallax feature not shared by existing schemes.
In addition to avoiding successive interference cancellation (SIC), each user independently demodulates its own bits on separate I and Q branches with calculations based on closed-form expressions.
Hence the receiver complexity is on par with that of orthogonal multiple access (OMA), which is much lower than that in other competing alternatives such as non-orthogonal multiple access (NOMA) and rate-splitting multiple access (RSMA).
We carry out system optimization and determine the achievable rate region.
Numerical results show that the proposed system has a larger rate region relative to other benchmark schemes with receivers not using SIC, and even achieves a comparable rate region to those benchmark schemes with SIC receivers.

\end{abstract}

\section{Introduction}
With the increasing demand for massive connectivity and high spectrum
efficiency (SE), next-generation multiple access (NGMA) has garnered
significant attention from researchers \cite{NGMA}.
In particular, the downlink multiple access design has evolved from conventional orthogonal multiple access (OMA) schemes, 
which leverage orthogonality in frequency, time and code domains, 
to embrace more efficient and flexible strategies.
For the single-input single-output (SISO) broadcast channel (BC), 
it has been proven that the combination of superposition coding (SC) and successive interference cancellation (SIC) achieves the capacity region\cite{CoverT_1972_BC, DTse_2005_fundamentals}.
In the case of the multiple-input single-output (MISO) Gaussian BC, 
dirty paper coding (DPC) has been shown to be the capacity-achieving strategy \cite{weingarten_2006_DPC}.
However, DPC suffers from the high complexity of the nonlinear processing at the
transmitter, which poses challenges to its practical implementations.

In contrast to DPC, space-division multiple access (SDMA) based on linear precoding has been widely adopted in multi-antenna systems owing to its lower transmitter complexity and near-capacity performance, particularly when user channels are nearly orthogonal with similar channel strengths \cite{GS_2003_SDMA_DOF,yoo_2006_ZF}.
In SDMA systems, inter-user interference is suppressed or even eliminated by skillful precoding, 
which enables low receiver complexity.  
However, the performance of SDMA decreases quickly in overloaded scenarios where user channels are highly correlated \cite{BC_2013_SDMA_condition}.

Inspired by the capacity-achieving performance in single-antenna systems, 
non-orthogonal multiple access (NOMA) relying on SC at the transmitter and SIC at the receiver is extended to the multi-antenna setup \cite{DLL_2017_NOMA_survey}.
Existing literature shows that multi-antenna NOMA achieves significant performance gains over OMA systems in an overloaded system that serves users with highly correlated channels and diverse channel strengths \cite{liu_2022_evolution, ding_2015_MIMO_NOMA}.

Over the past decade, a noticeable advancement in the area of multi-antenna multiple access has been the emergence of rate-splitting multiple access (RSMA)  \cite{MYJ22survey}. It unifies SDMA, NOMA and OMA into one general framework \cite{BClerckx19two_user}.
In conventional downlink RSMA systems, the transmitter splits and encodes all users' messages into a common stream and several private streams, which are then superposed and transmitted based on linear precoding.
SIC-based receivers are utilized to decode and remove the common stream from the received signal before decoding the intended private stream.
By partially treating interference as noise and partially decoding interference, 
RSMA shows numerous advantages, e.g. the achievable rate region close to that of DPC \cite{MYJ18simulation}, the robustness to imperfect channel state information at transmitter (CSIT) \cite{HJ_2016_TSP, DMB_16_TWC}, significant improvements in terms of spectral efficiency \cite{HJBC_2017_TWC}, energy efficiency \cite{MYJ_2018_EE} and degree of freedom (DoF) \cite{17CL_DoF}.
Furthermore, the advantages of RSMA are experimentally demonstrated on a software-defined radio testbed \cite{XL_BC_2024_RSMAProto}.

The research on NOMA and RSMA generally relies on SIC to decode and eliminate interference caused by message superposition.
On the one hand, SIC increases the computational complexity of the receiver, especially as the number of SIC layers grows in multi-user scenarios \cite{Bclerckx_2021_nomaefficient}.
On the other hand, it is difficult to implement perfect SIC in a practical system.
In particular, error propagation due to imperfect SIC can significantly undermine system performance \cite{Sena2020_EP}.
The limitations of SIC underscore the need to explore SIC-free multiple access schemes.

\subsection{Downlink Multiple Access with SIC-free Receivers}
In a downlink system with a single-antenna transmitter, SIC-free transmission schemes based on discrete constellations and lattice partitioning have been proposed in \cite{7416630} and \cite{8291591}, allowing single-user decoding and achieving rates within a constant gap from the capacity region.
In \cite{8681742}, the receiver employing modulo operation
instead of SIC is used in a multi-user scheduling setting. The complexity
reduction however comes with the cost of performance. A hierarchical quadrature amplitude modulation (H-QAM) 
based multiple access scheme, termed QAMA, has been proposed in
\cite{ZhZZ2023QAMA}, which assigns users separate bit streams from a tunable
hierarchical constellation. 
Even using simpler receiver processing, QAMA achieves a similar rate region to multi-user superposition transmission (MUST) in the 3GPP standard for 5G cellular systems with SIC receivers \cite{MUST}. 
In \cite{22CL_0SIC}, a single-antenna system design based on rate splitting
makes an effort to eliminate SIC where the common and private streams are carried in the in-phase and quadrature components of the transmitted signal.

In the systems with a multiple-antenna transmitter, there have been recent
works with SIC-free receivers. 
Taking advantage of the additional degrees of freedom (DoF) offered by the polarization domain, a dual-polarized massive multiple-input multiple-output (MIMO) RSMA system with three transmission strategies is proposed in \cite{23TWC_DPRSMA} to address practical issues caused by imperfect SIC. In particular, the first strategy avoids the use of SIC at the receiver.
Practical non-SIC receivers are considered in the 1-Layer RSMA system to assess
the performance of RSMA without SIC-based receivers
\cite{ZSB_24_SPAWC,ZSB_24_NSIC}. In these works, the superiority of RSMA over SDMA remains to some extent even without SIC at receivers. 
Numerical results also confirm performance gaps in terms of rate region between RSMA systems with SIC receivers and non-SIC receivers.

\subsection{Contributions}
\begin{figure}
    \centering
    \includegraphics[scale=0.9]{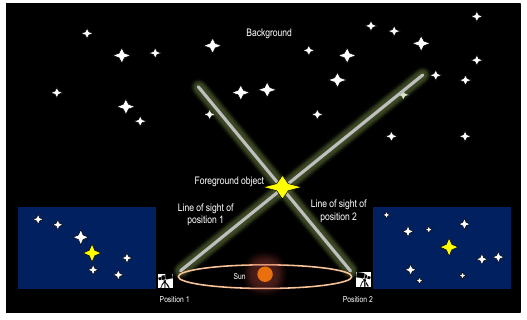}
    \caption[Stellar parallax in PxQAMA systems]{Stellar parallax refers to the apparent shift in a star’s position relative to the background as observed from the Earth, resulting in a composite pattern that varies with the Earth's motion. Similarly, an individual composite constellation is observed by each user due to different channels and precoding vectors in PxQAMA systems.}
    \label{fig:parallax}
\end{figure}

In this paper, we propose a novel downlink multiple access scheme, 
termed parallax QAMA (PxQAMA), 
where a multi-antenna transmitter simultaneously serves two single-antenna users.
In each channel use, coded bits from two users are split and mapped into one shared symbol and two private symbols, 
which are drawn from individual H-QAM constellations with different sizes and distance parameters.
Specifically, bit streams from the shared symbol are assigned to two users as in the QAMA framework, 
while each private symbol carries bit streams intended for a single user.
Carried on separate beams, the three symbols are superimposed and transmitted together.
The innovative designs of the proposed system are two-fold: 
\begin{itemize}
    \item The shared and private symbols are jointly Gray-mapped to ensure that the composite constellation remains Gray-coded.
    \item The orthogonality constraint and phase alignment constraint are imposed on the precoding design. 
\end{itemize}
At the receiver side, each user observes a Gray-coded H-QAM constellation, which appears similar to QAMA though it is actually formed by the over-the-air superposition of the shared and private symbols.
Moreover, the received constellations differ across users, as they have user-specific distance parameters shaped by their channels, resembling stellar parallax and inspiring the name, as shown in Fig.~\ref{fig:parallax}\footnote{The picture is adapted from https://www.space.com/30417-parallax.html}.
Note that the parallax effect is not available in either
QAMA or SDMA, where each user sees the same constellation as designed by the
transmitter.  This parallax effect is also not applicable to RSMA systems,
where the receiver does not see a regular constellation.

The contributions of this paper are summarized as follows.
\begin{itemize}
    \item The PxQAMA framework ensures low-complexity receiver processing.
    Each user only decodes the assigned bit streams in the shared symbol and the intended private symbol without consideration of other users.  
    Not only does it avoid the use of SIC, but it also enables soft demapping based on closed-form expressions on the I and Q branches independently, thanks to the Gray mapping. The receiver of PxQAMA remains as simple as in an OMA system, which is lower than other competing alternatives, including the SIC-free receivers presented for RSMA in \cite{ZSB_24_NSIC}.
    \item The PxQAMA system is optimized, with closed-form expressions on the precoder directions and phases. The optimization framework is presented to determine the achievable rate region. 
    \item Comparing the proposed system against mainstream multiple-access benchmarks, simulation results demonstrate that PxQAMA achieves superior performance with the simplest receiver.
    Unlike SDMA and NOMA, whose advantages are typically confined to specific channel conditions, PxQAMA performs consistently well in various scenarios.
    Compared to RSMA with SIC-free receivers, PxQAMA achieves a larger rate region with a simpler receiver. 
    Even compared to the RSMA with SIC-based receivers, PxQAMA achieves a comparable rate region in most cases.
\end{itemize}
While the focus of this paper is on a two-user system, the extension 
to multiple users is outlined, and an example of 
practical design with finite transmission modes is provided.
Thanks to the receiver simplicity and outstanding performance, PxQAMA is a 
promising downlink multiple access scheme for future wireless networks, 
especially when the receiver complexity is the main bottleneck 
for real time signal processing or on low-cost computation devices.

The rest of the paper is organized as follows. 
The system model and the transceiver designs 
are presented in Section \ref{sec:transceiver}. 
System optimization and the achievable rate region 
for the system are developed
in Section \ref{sec:rateregion}.
The achieved rate regions are compared in Section \ref{sec:result} and conclusions are collected in Section \ref{sec:conclusion}.

\textit{Notation:} Vectors and matrices are denoted by lower case and upper case of bold letters; the conjugate transpose of vector $\bbv$ is denoted by $\bbv^H$; $\Re\{\cdot\}$ and $\Im\{\cdot\}$ indicate the real and imaginary part of a complex number; $\|\cdot\|$ is the Euclidean norm; 
$\mathbb{E}\left[\cdot\right]$ refers to the statistic expectation.

\begin{figure*}[t]
    \centering
    \includegraphics[scale=0.45]{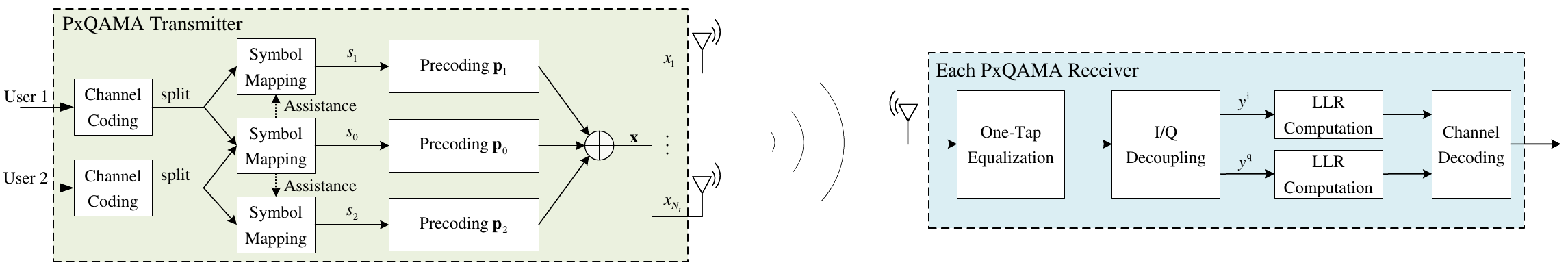}
    \caption{The transmitter and receiver diagram for the two-user PxQAMA system.}
    \label{fig: LSMA_transiver}
\end{figure*}

\section{System Description}
\label{sec:transceiver}

We consider a downlink system where a transmitter with 
$N_{\rm t}$ antennas serves $N_{\rm u}=2$ single-antenna users. 
Denoting the channel vector between the transmitter and user $u$ as
$\mathbf{h}_u\in \mathbb{C}^{N_{\rm t}\times 1}$, 
the received signal at user $u$ is 
\begin{equation}
	y_u = \bbh_u^H \bbx + w_u,\ \ u\in\{1,2\},
\end{equation}
where $\bbx$ is the transmitted signal vector 
and $w_u$ is the additive white
Gaussian noise (AWGN) at user $u$ that follows 
circularly symmetric complex Gaussian 
distribution with zero mean and variance $\sigma^2$.

Fig.~\ref{fig: LSMA_transiver} depicts the proposed system diagram.  As in an
OMA system, each user has a separate channel encoder and channel decoder, and
the novel system component is on the modulator for downlink multiple access. 
As shown in Fig.~\ref{fig: LSMA_transiver}, the coded bits from two users 
are split and mapped into one shared symbol $s_0$ and two private symbols $s_1,s_2$.
The shared symbol carries bit streams assigned to both users, as in the QAMA framework, while each private symbol carries bit streams intended for a single user.
At each channel use, the transmitted signal vector $\bbx$ is 
\begin{equation}
    \bbx = \bbp_0 s_0+\bbp_1s_1 +\bbp_2 s_2,
    \label{ori_transmit_x}
\end{equation}
where $\bbp_i \in \mathbb{C}^{N_t\times 1}$, $i\in\{0,1,2\}$, is the precoding vector corresponding to $s_i$.

The constellations for $s_0$, $s_1$ and $s_2$ are drawn from 
hierarchical $M$-ary QAM (H-$M$QAM) constellations with
different sizes and distance parameters. 
For symbol $s_i$, the H-QAM constellation can be decomposed 
into a $2^{m_i}$-PAM on the I branch and a $2^{n_i}$-PAM on the Q branch,
and hence carries $m_i+n_i$ bits per symbol \cite{1159785Recursive, 1299057BER4GQAM}.
Define the collection of all the bits corresponding to symbol $s_i$ as:
\begin{equation}
	\mathcal{S}_{i} := 
	\left\{b^{\rm i}_{i,\kappa} \right\}_{\kappa=1}^{m_i} \cup 
	\left\{b^{\rm q}_{i,l} \right\}_{l =1}^{n_i},
\end{equation}
and the collection of the distance parameters as: 
\begin{equation}
\mathcal{D}_{i} := 
\left\{d^{\rm i}_{i,\kappa}\right\}_{\kappa=1}^{m_i} \cup
\left\{d^{\rm q}_{i,l} \right\}_{l =1}^{n_i}.
\label{eq:distanceHQAM}
\end{equation}
Each constellation is normalized to have unit energy that requires: 
\begin{equation}
\sum_{\kappa=1}^{m_{i}}(d^{\rm i}_{i, \kappa})^2+\sum_{l=1}^{n_{i}}(d^{\rm q}_{i, l})^2 =1.
        \label{eq:Es}
\end{equation}
On each branch, the distance parameter corresponding to the lower-layer bit should be at least twice that of the higher-layer bit, i.e. $d^{\rm i}_{i, \kappa}\ge 2d^{\rm i}_{i, \kappa+1},d^{\rm q}_{i, l}\ge 2d^{\rm q}_{i,l+1}$, to preserve the correct ordering of the constellation.

With perfect channel state information (CSI) at the transmitter, the system
aims to optimize the data rates for both users while maintaining simple
processing at each receiver. We next present the framework for the transmitter
and receiver, while deferring the system optimization to Section
\ref{sec:rateregion}.

\subsection{Transmitter Design}
\label{sec: transmitter}

The PxQAMA framework judiciously designs the symbol 
mappers and the precoders at the transmitter.

\subsubsection{Joint Symbol Mapping}

The shared symbol $s_0$ adopts conventional Gray mapping for H-QAM \cite{1159785Recursive, 1299057BER4GQAM}. 
As clarified in Appendix \ref{sec:insight},
the information symbol in Gray mapping is related to 
the information bits as
\begin{equation}
\left\{
\begin{aligned}
    \Re \{s_0\}&=\sum_{\kappa=1}^{m_{0}}(-1)^{(1+\sum_{j=1}^{\kappa}b^{\rm i}_{0, j})}d^{\rm i}_{0, \kappa}\\
    \Im \{s_0\}&=\sum_{l=1}^{n_{0}}(-1)^{(1+\sum_{j=1}^{l}b^{\rm q}_{0,j})}d^{\rm q}_{0,l}.
    \label{eq:s0}
\end{aligned}
\right.
\end{equation}
For private symbols $s_1$ and $s_2$, we propose a novel mapping rule as 
\begin{equation}
\left\{
    \begin{aligned}
        \Re\{s_{i}\}&=(-1)^{(1+\sum_{\kappa=1}^{m_{0}}b^{\rm i}_{0,\kappa})}
        \sum_{\kappa=1}^{m_{i}}(-1)^{(\sum_{j=1}^{\kappa}b^{\rm i}_{i, j})}
        d^{\rm i}_{i, \kappa},\\
        \Im\{s_{i}\}&=
        (-1)^{(1+\sum_{l=1}^{n_{0}}b^{\rm q}_{0,l})}
        \sum_{l=1}^{n_{i}}(-1)^{(\sum_{j=1}^{l}b^{\rm q}_{i,j})}
        d^{\rm q}_{i, l}.
    \end{aligned}
    \right.
    \label{eq:sk}
\end{equation}
Apparently, the Gray mapping for $s_1, s_2$ is dependent on the 
bits in $s_0$. This mapping rule allows the following effects.
\begin{itemize} 
	\item $s_0+c_1 s_1$ has Gray mapping 
		for the collection of bits in $\calS_0$ and $\calS_1$,
		where $c_1$ is a suitable positive constant.
	\item $s_0+c_2 s_2$ has Gray mapping 
		for the collection of bits in $\calS_0$ and $\calS_2$
		where $c_2$ is a suitable positive constant.
\end{itemize} 
An illustration of this effect is provided in Appendix \ref{sec:insight}.

\medskip
\subsubsection{Constraints on Precoding Vectors} 

We adopt the following constraints on the precoding vectors. 
\begin{itemize} 
	\item Direction control for $\bbp_1$ and $\bbp_2$:
\begin{equation}
    \bbh_1^H\bbp_2=0, \quad \bbh_2^H\bbp_1=0.
    \label{eq: orthog_const}
\end{equation}

\item Phase alignment for $\bbp_1$ and $\bbp_2$:
\begin{equation}
    \left\{
    \begin{aligned}
	    e^{j\angle(\bbh_1^H\bbp_1)} &= e^{j\angle(\bbh_1^H\bbp_0)} = e^{j\Phi_1},\\
	    e^{j\angle(\bbh_2^H\bbp_2)} &= e^{j\angle(\bbh_2^H\bbp_0)} = e^{j\Phi_2},
    \end{aligned}\right.
    \label{eq:phase_const}
\end{equation}
where $\angle(\cdot)$ is the phase of a complex number.
\end{itemize}
\vspace{0.5cm}
The direction and phase of $\bbp_0$ are left for optimization.

\subsection{The Parallax Effect}

Subject to the direction control in \eqref{eq: orthog_const}, the 
received signal for user $u$, $u=1,2$, can be expressed as
\begin{equation}
    y_u = \bbh_u^H\bbp_0s_0 + \bbh_u^H\bbp_us_u+w_u.
    \label{recev_y}
\end{equation} 
For brevity, define the following variables: 
\begin{equation} 
G_u = \sqrt{|\bbh_u^H\bbp_0|^2+|\bbh_u^H\bbp_u|^2},
\end{equation} 
\begin{equation} 
	\beta_{u,0} = \frac{|\bbh_u^H\bbp_0|} {G_u}, \quad 
	\beta_{u,u} = \frac{|\bbh_u^H\bbp_u|} {G_u} .
\end{equation}
With the phases aligned in \eqref{eq:phase_const}, 
one can simplify \eqref{recev_y} as
\begin{equation}
y_u = e^{j\Phi_u} G_u 
	\left( \beta_{u,0}s_0 + \beta_{u,u} s_u \right) +w_u.
\label{recev_y2}
\end{equation} 
Now introduce a composite symbol formed by $s_0$ and $s_u$ as:
\begin{equation}
	\tilde s_u = \beta_{u,0} s_0 + \beta_{u,u} s_u.
\label{recev_y3}
\end{equation} 
The input-output relationship in \eqref{recev_y2} is equivalent to
\begin{equation}
	y_u = e^{j\Phi_u} G_u \tilde s_u +w_u.
\label{eq:io_equ}
\end{equation} 

Let us inspect the symbol $\tilde s_u$. It can be viewed 
as a symbol drawn from a H-QAM constellation 
with $m_0+m_u$ bits on the I-branch and $n_0+n_u$ bits 
on the Q-branch. Indeed, let us define the equivalent bits and distances as 
\begin{align}
    \tilde{b}^{\rm i}_{u,\kappa} &=
    \begin{cases}
	    b^{\rm i}_{0,\kappa}, & \text{if $\kappa\leq m_0$}\\
	    b^{\rm i}_{u,\kappa-m_0}, & \text{if $\kappa>m_0$}
    \end{cases}
    \label{rece_bi}\\
    \tilde{b}^{\rm q}_{u,l} &=
    \begin{cases}
	    b^{\rm q}_{0,l}, & \text{if $l\leq n_0$}\\
	    b^{\rm q}_{u,l-n_0}, & \text{if $l>n_0$}
    \end{cases}
    \label{rece_bq}
\end{align}
\begin{align}
    \tilde{d}^{\rm i}_{u,\kappa} &=
    \begin{cases}
	    \beta_{u,0}d^{\rm i}_{0,\kappa}, & \text{if $\kappa\leq m_0$}\\
	    \beta_{u,u} d^{\rm i}_{u,\kappa-m_0}, & \text{if $\kappa>m_0$}
    \end{cases}
    \label{rece_d1}\\
    \tilde{d}^{\rm q}_{u,l} &=
    \begin{cases}
	    \beta_{u,0} d^{\rm q}_{0,l}, & \text{if $l\leq n_0$}\\
	    \beta_{u,u} d^{\rm q}_{u,l-n_0}, & \text{if $l>n_0$}.
    \end{cases}
    \label{rece_d2}
\end{align}
By expressing $\tilde s_u$ as:  
\begin{equation}
\left\{
\begin{aligned}
        \Re\{\tilde s_{u}\}&=
        \sum_{\kappa=1}^{m_0+m_{u}}(-1)^{(1+\sum_{{j}=1}^{\kappa}
	\tilde b^{\rm i}_{u, j})}
        \tilde d^{\rm i}_{u, \kappa},\\
        \Im\{\tilde s_{u}\}&=
        \sum_{l=1}^{n_0+n_{u}}(-1)^{(1+\sum_{{j}=1}^{l}
	\tilde b^{\rm q}_{u,{j}})} \tilde d^{\rm q}_{u, l},
    \end{aligned}
    \right.
    \label{eq:su}
\end{equation}
one can see that Gray mapping from bits to symbols is maintained. 
Also, the constellation for $\tilde s_u$ has unit energy as 
the distances in \eqref{rece_d1} and \eqref{rece_d2} satisfy the constraint in \eqref{eq:Es}. 

We emphasize that the equivalent constellations for user 1 and user 2 are
different as they have different distance parameters shaped by their channels.
Inspired by the terminology from astronomy (Fig.~\ref{fig:parallax}), we refer to
this as the parallax effect of the proposed system design. Note that this
parallel effect is absent in QAMA and SDMA systems, where each user sees the same constellation as designed by the
transmitter. 
This parallax effect is also not applicable to RSMA systems, where the receiver does not see a regular constellation.

\begin{figure*}[t]
    \centering
    \includegraphics[scale=0.78]{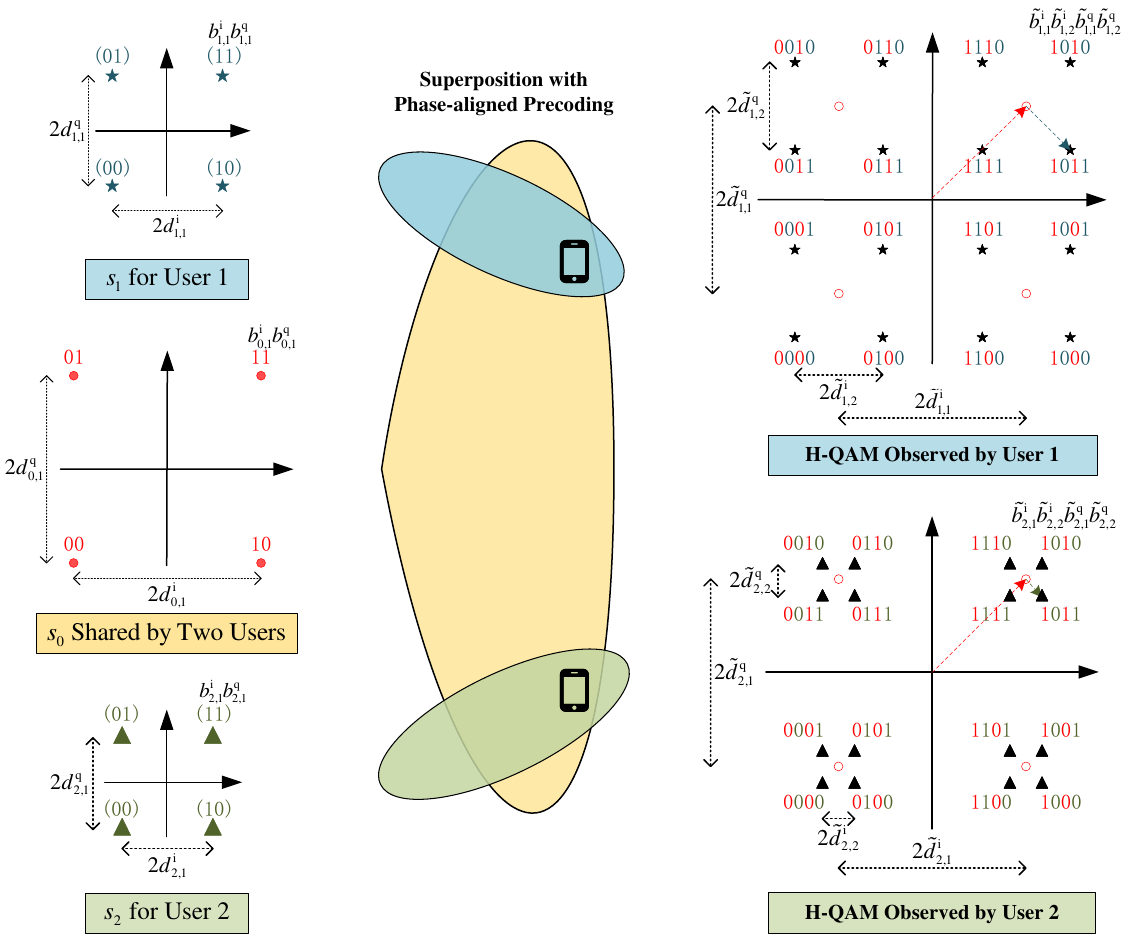}
    \caption{Illustration of the system design 
	with $s_0,s_1,s_2$ drawn from QPSK symbols, i.e., $m_i=n_i=1,\forall i$.
	Two receivers observe Gray-coded H-16QAM 
	constellations with different distance parameters. 
    }
    \label{fig: overalldemo}
\end{figure*}

Fig.~\ref{fig: overalldemo} provides an illustration with $s_0,s_1,s_2$ drawn
from QPSK constellations, while each receiver observes a different H-16QAM. 

\textit{Remark:}
A similar layered constellation structure could be observed at the receiver of layered division multiplexing (LDM) systems in Advanced Television Systems Committee (ATSC) 3.0 standard \cite{ATSC3dot0}, 
where the core and extended layers are superimposed to reach receivers with different reception qualities \cite{ZL_2016_LDM}. 
However, in PxQAMA, the superposition of shared and private symbols occurs over the air, resulting in each user observing a distinct constellation. 
In contrast, the layered structure in LDM is formed at the transmitter, where all users receive an identical constellation consisting of the same core and extended layers.

\subsection{Simple Receiver Processing}
\label{sec:receiver_complexity}

Based on \eqref{eq:io_equ}, each user demodulates the bits contained in symbol
$\tilde s_u$ from the received signal $y_u$. 
Since different receivers follow the same steps, 
we drop the user index and simplify \eqref{eq:io_equ} as 
\begin{equation}
y = e^{j\Phi} G \tilde s + w,
\end{equation}
where $e^{j\Phi} G$ is the equivalent complex channel gain and should be determined before the receiver processing.

Regardless of how bit streams are layered and assigned, the receiver process of
each user is as simple as the single-user receiver in an OMA system with a uniform
QAM constellation. With more details available in \cite{ZhZZ2023QAMA},
we summarize the main steps as follows. 
\begin{itemize}

\item {\bf Step 1}: 
	One-tap equalization and I/Q decoupling are first performed on the received signal. 

    \begin{equation}
    \tilde{y}^{\rm i} := \Re\left\{\frac{e^{-j\Phi}y}{G}\right\} = 
	    \tilde{s}^{\rm i}+ \tilde w^{\rm i} ,\\
        \label{eq:equalized_siganl_i}
    \end{equation}
    \begin{equation}
    \tilde{y}^{\rm q} := \Im\left\{\frac{e^{-j\Phi}y}{G}\right\} = 
	    \tilde{s}^{\rm q}+ \tilde w^{\rm q},
        \label{eq:equalized_siganl_q}
    \end{equation}
where $\tilde w^{\rm i}$ and $\tilde w^{\rm q}$ are the resulting noises.

    \item {\bf Step 2}: Based on the distance parameters of the received constellation, each receiver calculates the log-likelihood ratio (LLR) of its own assigned bits. 
    The real and imaginary components are processed separately in the same way.
    Taking the I branch as an example, the LLR for the $\kappa$-th bit $\tilde b_{\kappa}^{\rm i}$ is defined as
\begin{equation}
	\Lambda(\tilde b_{\kappa}^{\rm i} ) = \ln 
	\frac{ f(\tilde{y}^{\rm i} 
	|\tilde b_{\kappa}^{\rm i} =1) } { f(\tilde{y}^{\rm i}|\tilde b_{\kappa}^{\rm i}=0)},
    \label{eq: exact_llr}
\end{equation}
where $f(\tilde{y}^{\rm i}|\tilde{b}_{\kappa}^{\rm i})$ denotes the likelihood function of the measurement $\tilde{y}^{\rm i}$ given $\tilde b_{\kappa}^{\rm i}$.
With the following definition
\begin{equation}
    z_{\kappa}^{\rm i} = \min_{\tilde{s} \in \calS(\tilde b_{\kappa}^{\rm i}=0)} 
	{\frac 14|\tilde{y}^{\rm i}-\tilde{s}^{\rm i}|^2}
	-	\min_{\tilde{s} \in \calS(\tilde b_{\kappa}^{\rm i}=1)} 
	{\frac 14|\tilde{y}^{\rm i}-\tilde{s}^{\rm i}|^2},
 \label{eq:zk_cal}
\end{equation}
		the bit LLR is approximately by \cite{ZhZZ2023QAMA}:
\begin{equation}
 \Lambda(\tilde b_{\kappa}^{\rm i}) \approx \frac{4|G|^2}{\sigma^2} z_{\kappa}^{\rm i}.
\end{equation}

Note that a piecewise linear mapping operation can replace the dual-$\min$ operation
		in \eqref{eq:zk_cal} for further complexity reduction
		\cite{ZhZZ2023QAMA}. 
Taking the I branch of H-64QAM as an example, i.e., a hierarchical 8-PAM constellation, 
the approximated bit LLRs can be obtained as \cite{ZhZZ2023QAMA}
\begin{align}
\label{8PAMz1}
        z^{\rm i}_{1} &= 
  \begin{cases}
            \tilde{y}^{\rm i}(\tilde{d}^{\rm i}_1-\tilde{d}^{\rm i}_2-\tilde{d}^{\rm i}_3),\ &|\tilde{y}^{\rm i}|\in[0,\tilde{d}^{\rm i}_1-\tilde{d}^{\rm i}_2)\\
            (\tilde{d}^{\rm i}_1-\tilde{d}^{\rm i}_2)(\tilde{y}^{\rm i}-\text{sign}(\tilde{y}^{\rm i})\tilde{d}^{\rm i}_3),\ &|\tilde{y}^{\rm i}|\in[\tilde{d}^{\rm i}_1-\tilde{d}^{\rm i}_2,\tilde{d}^{\rm i}_1)\\
            (\tilde{d}^{\rm i}_1-\tilde{d}^{\rm i}_3)(\tilde{y}^{\rm i}-\text{sign}(\tilde{y}^{\rm i})\tilde{d}^{\rm i}_2),\ &|\tilde{y}^{\rm i}|\in[\tilde{d}^{\rm i}_1,\tilde{d}^{\rm i}_1+\tilde{d}^{\rm i}_2)\\
            \tilde{d}^{\rm i}_1(\tilde{y}^{\rm i}-\text{sign}(\tilde{y}^{\rm i})(\tilde{d}^{\rm i}_2+\tilde{d}^{\rm i}_3)),\ &|\tilde{y}^{\rm i}|\in[\tilde{d}^{\rm i}_1+\tilde{d}^{\rm i}_2,\infty)\\
  \end{cases}
        \\
        \label{8PAMz2}
	z^{\rm i}_2 &= 
  \begin{cases}
        \tilde{d}^{\rm i}_2 (\tilde{d}^{\rm i}_1 - \tilde{d}^{\rm i}_3-|\tilde{y}^{\rm i}|),&|\tilde{y}^{\rm i}|\in [0, \tilde{d}^{\rm i}_1-\tilde{d}^{\rm i}_2) \\
        (\tilde{d}^{\rm i}_2 - \tilde{d}^{\rm i}_3) (\tilde{d}^{\rm i}_1-|\tilde{y}^{\rm i}|),&|\tilde{y}^{\rm i}|\in [\tilde{d}^{\rm i}_1-\tilde{d}^{\rm i}_2, \tilde{d}^{\rm i}_1+\tilde{d}^{\rm i}_2) \\
        \tilde{d}^{\rm i}_2 (\tilde{d}^{\rm i}_1 + \tilde{d}^{\rm i}_3-|\tilde{y}^{\rm i}|),&|\tilde{y}^{\rm i}|\in [\tilde{d}^{\rm i}_1+\tilde{d}^{\rm i}_2, \infty) \\
  \end{cases}
	\\
    \label{8PAMz3}
	z^{\rm i}_3 &= \tilde{d}^{\rm i}_3 (\tilde{d}^{\rm i}_2 - \big|\tilde{d}^{\rm i}_1-|\tilde{y}^{\rm i}|\big|).
\end{align}

\item {\bf Step 3}: Each user forwards the LLRs of its assigned bits 
		into a channel decoder to recover the original message.  

\end{itemize}
\vspace{0.5cm}

No SIC is needed at the receiver. Further, we have  
the following remarks on complexity. 
\begin{itemize} 

	\item Each receiver has the same complexity as that in an OMA system
		with uniform QAM of the same constellation size
		$2^{m_0+m_u}\cdot 2^{m_0+n_u}$. Indeed, the same LLR
		computations can be used in both systems with the only
		difference on the values of the distance parameters.

	\item In all the simulation results, the constellation sizes for all
		receivers in PxQAMA are no larger than H-64QAM.  Hence, the LLRs
		for three bits on the I branch and another
		three bits on the Q branch are computed based on the
		closed-form expressions in \eqref{8PAMz1}-\eqref{8PAMz3}.

\end{itemize}

\subsection{Two Special Cases}
\label{sec:specical_designs}
\begin{figure}[t]
    \centering
    \subfloat[{\footnotesize QAMA-BF}]{\includegraphics[scale=0.4]{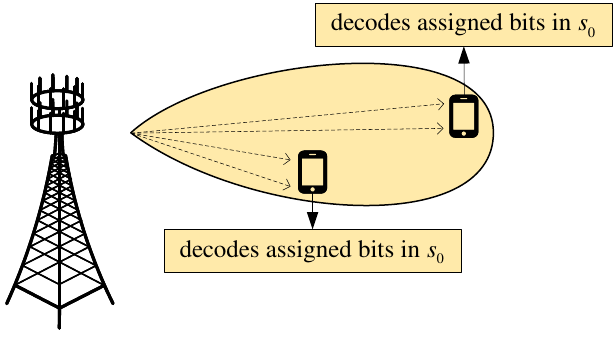}
    \label{spedemo_subfig_1}}
    \subfloat[{\footnotesize SDMA}]{\includegraphics[scale=0.4]{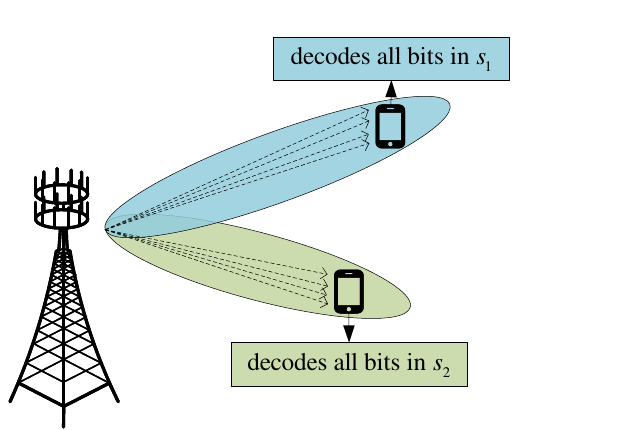}
    \label{spedemo_subfig_2}}
	\caption{The illustration of two special cases:
	(a) QAMA-BF and (b) SDMA, where 
	the dashed lines represent the 
	four bit streams considered in each beam.}
\label{fig:demo_specialcases}
\end{figure}

The general PxQAMA framework includes two important special cases as shown in Fig. \ref{fig:demo_specialcases}.
\begin{itemize} 
\item {\bf QAMA with beamforming}: In this case, only one beam is formed with $\bbx = \bbp_0 s_0$, and the bits from two users are carried via the shared symbol $s_0$, i.e., $m_1=m_2=n_1=n_2=0$.
Each user decodes the assigned bit streams in the received signal:
\begin{equation} 
	y_u = \bbh_u^H\bbp_0s_0 + w_u.  
\end{equation}
This special case corresponds to the combination of QAMA with beamforming (QAMA-BF). 
Intuitively, this special configuration is useful
when the channels from two users are highly correlated.

\item {\bf SDMA}: This configuration with no shared symbol, i.e., $m_0=n_0=0$, is reduced to SDMA 
	with $\bbx=\bbp_1 s_1+\bbp_2 s_2$. With the directional control in 
\eqref{eq: orthog_const}, one obtains
\begin{equation}
    y_u = \bbh_u^H \bbp_{u} s_{u} + w_u.
    \label{eq:SDMAy}
\end{equation}
Intuitively, this special configuration is useful when 
the channels from two users are nearly orthogonal. 

\end{itemize} 

\subsection{Extension to multi-user systems}
We outline the extension of the PxQAMA framework to cases with $N_{\rm u}>2$ users.
Similarly, coded bits from all users are split and mapped into one shared symbol $s_0$ and $N_{\rm u}$ private symbols $s_i, i\in\{1,...,N_{\rm u}\}$.
Bit streams from the shared symbol are assigned to different users and each private symbol is intended for a single user.
Joint symbol mapping in \eqref{eq:s0} and \eqref{eq:sk} is adopted to ensure that the composite constellation of $s_0$ and $s_{i}$ remains Gary-coded.
Then the transmitted signal vector at each channel use becomes
\begin{equation}
    \bbx=\bbp_0s_0+\sum_{i=1}^{N_{\rm u}}\bbp_is_i.
\end{equation}
The directions of private symbol precoders are constrained as 
\begin{equation}
    \bbh_u^H\bbp_i=0, \forall u\neq i.
\end{equation}
Hence, $\bbp_i$ is drawn from the null space of  the channel vectors  $\{\bbh_1, \dots, \bbh_{i-1}, \bbh_{i+1}, \dots, \bbh_{N_{\rm u}}\}$.
The phase alignment constraint for private symbol precoders becomes
\begin{equation}
    e^{j\angle(\bbh_i^H\bbp_i)} = e^{j\angle(\bbh_i^H\bbp_0)} = e^{j\Phi_i}, \forall i.
\end{equation}

As a result, the parallax effect remains in the multi-user cases, where each user observes a Gray-coded H-QAM constellation with individual distance parameters.

\subsection{PxQAMA vs. One-layer RSMA}

Although sharing the same expression for the transmitted vector in
\eqref{ori_transmit_x}, PxQAMA and one-layer RSMA \cite{MYJ22survey} are
distinct from each other. 

\begin{itemize} 

	\item The ``message splitter'' in RSMA is done on the information bits
		before channel coding.  For a two-user system, 	there are three
		channel encoders at the transmitter side, one for the common
		stream, and one for each user's private stream.  Each receiver
		deploys two channel decoders, one for the common stream, and
		one for its own private stream.  In total, there are three
		encoders at the transmitter and four decoders at the receivers
		for a two-user one-layer RSMA system.  The split and joint modulation
		in PxQAMA is done on the bits after channel coding. There are
		two encoders and two decoders in the PxQAMA system.

	\item In RSMA, the common stream needs to be decoded successfully by all
		users. As a result, the data rate of the common stream is
		limited by the user with the worst channel quality. 
            In PxQAMA, the shared symbol carries bit streams assigned to different users. Hence each receiver only demodulates its own bits, without any
		operation on the bits from the other user.

	\item  RSMA can adopt SIC-free receivers to avoid the drawbacks of SIC
		receivers \cite{ZSB_24_NSIC}. However, the complexity of the SIC-free
            receivers are still higher than the simple receivers for PxQAMA. This is
		because each user in the PxQAMA system observes a hierarchical
		QAM constellation with Gray mapping, which can be efficiently
		processed at the I and Q branches with simple LLR computation.
		RSMA adopts conventional constellations for the common and
		private data streams and the composite constellation at each
		receiver does not have a well-defined structure with
		Gray-mapping.
\end{itemize} 

\section{System Optimization}
\label{sec:rateregion}
In this section, our objective is to determine the achievable rate region of the proposed system.
To that end, we determine the closed-form solution to the precoder direction and phase shift,
followed by the information rate calculation and the optimization framework for the proposed system.

\subsection{Precoder Design}

For convenience, we separate the norm of a vector from its direction by
introducing unit-norm vectors as
\begin{equation}
    \left\{
    \begin{aligned}
        \bbh_u&=|\bbh_u|\tilde\bbh_u=\lambda_u\tilde\bbh_u,\ \ u=1,2,\\
        \bbp_i&=|\bbp_i|\tilde\bbp_i=\alpha_i\tilde\bbp_i,\ \ i=0,1,2,
    \end{aligned}\right.
\end{equation}
where the vectors marked with tilde have unit norms.
The variables $\lambda_u$ and $\alpha_i$ respectively denote 
the norm of the channel and the norm of the precoding vector, respectively.
Define the complex correlation coefficient $\rho$ as
\begin{equation}
     \rho=|\rho| e^{j\angle \rho}=\tilde\bbh_1^H\tilde\bbh_2,
\end{equation}
where $\angle \rho \in [-\pi, \pi]$ is called the vectors' pseudo-angle \cite{Scharnhorst01}. 
Moreover, the Hermitian angle $\Theta$ between the normalized channel vectors $\tilde\bbh_1$ and $\tilde\bbh_2$ is defined as  \cite{Conway&Hardin&Sloane:96Math}
\begin{equation}
	\Theta = \text{arccos}(|\rho|),
    \label{theta_def}
\end{equation}
and the chordal distance between $\tilde\bbh_1$ and $\tilde\bbh_2$
is  \cite{Conway&Hardin&Sloane:96Math}
\begin{equation}
	\sin(\Theta)=\sqrt{1-|\rho|^2}.
\end{equation}
Through the Gram-Schmidt orthogonalization process, one can find two basic directions that cover the space spanned by $[\bbh_1, \bbh_2]$ as
\begin{equation}
\left\{
    \begin{aligned}
	\tilde\bbq_1 &= \tilde\bbh_1, \\
	\tilde\bbq_2 &= \frac{\tilde \bbh_2 - \rho \tilde \bbh_1}{\sqrt{1-|\rho|^2}}.
    \label{GS_q}
    \end{aligned}
\right.
\end{equation}

\subsubsection{Key Parameter for Shared Symbol Precoder Direction}
For the shared symbol, the optimal precoding vector $\tilde\bbp_0$ shall be placed in the range space of
$\tilde\bbh_1$ and $\tilde\bbh_2$. 
Otherwise, energy is lost without benefiting any user.
We can hence set the precoding direction of the shared symbol as a linear combination of $\tilde\bbq_1$ and $\tilde\bbq_2$ as
\begin{equation}
	\tilde\bbp_0 = [\tilde\bbq_1, \tilde\bbq_2]
	\begin{bmatrix}
		\cos(\theta_0)\\
		e^{j\phi_0}\sin(\theta_0)
	\end{bmatrix},
 \label{pre_bf}
\end{equation}
where $\phi_0$ and $\theta_0$ are tunable parameters. By combining \eqref{GS_q} and \eqref{pre_bf}, one can verify that 
\begin{equation}
\left\{
    \begin{aligned}
        |\tilde\bbh_1^H \tilde\bbp_0| &= \cos(\theta_0),\\
        |\tilde\bbh_2^H \tilde\bbp_0| &=  |\rho^* \cos(\theta_0)+\sqrt{1-|\rho|^2} \sin(\theta_0) e^{j\phi_0}|,
    \end{aligned}
\right.
\label{eq:witheta}
\end{equation}
To maximize the right side of the second equation in \eqref{eq:witheta}, the two complex terms that are added together must have aligned phases. 
So we can first determine the optimal $\phi_0 = -\angle\rho$. Then with
\eqref{theta_def}, we can verify that $|\tilde\bbh_2^H \tilde\bbp_0|$ reaches
its maximum value of $|\cos(\Theta-\theta_0)|$.
Hence, we express the optimized beam direction corresponding to the shared symbol as
\begin{align}
		\tilde\bbp_0 &= 
		\cos(\theta_0)\tilde\bbq_1 +e^{-j\angle\rho} \sin(\theta_0)\tilde \bbq_2
		\notag\\
		&=
		\frac{\sin(\Theta-\theta_0)}{\sin(\Theta)} \tilde \bbh_1
	+
	\frac{\sin(\theta_0)}{\sin(\Theta)} e^{-j\angle\rho}\tilde \bbh_2,
\label{eq:opt_qamap_2users}
\end{align}
which has only one tunable parameter $\theta_0$. Substituting \eqref{eq:opt_qamap_2users} into \eqref{eq:witheta}, we have
\begin{equation}
    \left\{
    \begin{aligned}
        \tilde\bbh_1^H \tilde\bbp_0 &= \cos(\theta_0),\\
	    \tilde\bbh_2^H \tilde\bbp_0 &= \cos(\Theta-\theta_0)e^{-j\angle\rho}.
    \end{aligned}
    \right.
    \label{eq:QAMAsnr_2users}
\end{equation}
It reveals that the beam direction determined by $\theta_0$ is a trade-off between two users' equivalent channel gains of the shared symbol.
\subsubsection{Closed-Form Solution to Private Symbol Precoders}
Given the shared symbol precoder $\bbp_0$ in \eqref{eq:opt_qamap_2users}, we further determine the precoding vector for two private symbols, whose direction and phase shift are constrained by \eqref{eq: orthog_const} and \eqref{eq:phase_const}.

Choosing the directions in the range space of $\tilde\bbh_1$ and $\tilde\bbh_2$, we can explicitly express the directions satisfying the orthogonality constraint as
\begin{equation}
    \left\{
    \begin{aligned}
        \tilde\bbp_1 &= \frac{e^{j\phi_1}}{\sqrt{1-|\rho|^2}}(\tilde\bbh_1-\rho^*\tilde\bbh_2),\\
        \tilde\bbp_2 &= \frac{e^{j\phi_2}}{\sqrt{1-|\rho|^2}}(\tilde\bbh_2-\rho\tilde\bbh_1),\\
    \end{aligned}\right.
    \label{eq: precoder_e0}
\end{equation}
where $\phi_1$ and $\phi_2$ are tunable phase parameters for $\tilde\bbp_1$ and $\tilde\bbp_2$.
Hence we have $\angle(\tilde\bbh_1^H\tilde\bbp_1) = \phi_1, \angle(\tilde\bbh_2^H\tilde\bbp_2) = \phi_2$.
The phase alignment constraint in \eqref{eq:phase_const} could be satisfied by setting $\phi_1 = 0, \phi_2 = -\angle\rho$.
Hence we have the closed-form solution to the extended symbol precoder direction as
\begin{equation}
    \left\{
    \begin{aligned}
        \tilde\bbp_1 &= \frac{1}{\sqrt{1-|\rho|^2}}(\tilde\bbh_1-\rho^*\tilde\bbh_2),\\
        \tilde\bbp_2 &= \frac{e^{-j\angle\rho}}{\sqrt{1-|\rho|^2}}(\tilde\bbh_2-\rho\tilde\bbh_1).\\
    \end{aligned}\right.
    \label{eq: precoder_e}
\end{equation}

To summarize, we decompose the precoder design into power and direction components, focusing on the latter in this subsection.
The shared symbol precoder direction is controlled by $\theta_0$, which balances the equivalent channel gains of the two users.
For the private symbol, the precoding vectors have a closed-form solution as in \eqref{eq: precoder_e} due to the orthogonality and phase alignment constraints.

\subsection{Information Rates for the Bit Streams}

Substituting precoder designs in \eqref{eq:opt_qamap_2users} and \eqref{eq:
precoder_e} into the received signal model in \eqref{eq:io_equ}, we have
\begin{equation}
    \begin{aligned}
        y_1 &= G_1 \tilde s_1 +w_1,\\
        y_2 &= e^{-j\angle\rho}G_2 \tilde s_2+ w_2.
    \end{aligned}
    \label{eq:re_y_simp}
\end{equation}
where the modulus of equivalent channel gains are
\begin{equation} 
\begin{aligned}
	G_1& =\sqrt{[\lambda_1\alpha_0\cos(\theta_0)]^2 +[\lambda_1\alpha_1\sin(\Theta)]^2} \\ 
	G_2 &=\sqrt{[\lambda_2\alpha_0\cos(\Theta-\theta_0)]^2 +[\lambda_2\alpha_2\sin(\Theta)]^2} 
\end{aligned}
\end{equation}

The distance parameters in 
\eqref{rece_d1} and \eqref{rece_d2} for the constellations of $\tilde s_1$ and $\tilde s_2$ on the I branch
can be simplified to 
\begin{equation}
    \begin{aligned}
        \tilde{d}^{\rm i}_{1,\kappa} &=\begin{cases}            \displaystyle\frac{\lambda_1\alpha_0\cos(\theta_0)}{G_1}d^{\rm i}_{0,\kappa} & \text{if $\kappa\leq m_0$},\\
            \displaystyle\frac{\lambda_1\alpha_1\sin(\Theta)}{G_1}d^{\rm i}_{1,\kappa-m_0} & \text{if $\kappa> m_0$},
        \end{cases}\\
        \tilde{d}^{\rm i}_{2,\kappa}&=\begin{cases}
            \displaystyle\frac{\lambda_2\alpha_0\cos(\Theta-\theta_0)}{G_2}d^{\rm i}_{0,\kappa} & \text{if $\kappa\leq m_0$},\\
            \displaystyle\frac{\lambda_2\alpha_2\sin(\Theta)}{G_2}d^{\rm i}_{2,\kappa-m_0} & \text{if $\kappa> m_0$},
        \end{cases}
    \end{aligned}
    \label{rece_d_simp}
\end{equation}
The expressions on the Q branch are similar and hence omitted for brevity. 

Based on the equivalent channel and the known distance parameters for the
constellation, the information rate of bit streams can be determined. 
Specifically, based on the received symbol $y_u$, 
the LLRs for the bits in $\tilde s_u$, i.e., 
$\{ \tilde b_{u,\kappa}^{\rm i}, \tilde b_{u,l}^{\rm q}\}$ are generated as 
$\{ z_{u,\kappa}^{\rm i}, z_{u,l}^{\rm q}\}$.
The mutual information between each bit and the LLR can be found via the
reduction of the uncertainty when 
conditioned on the bit value, as detailed in \cite{ZhZZ2023QAMA}. 
For user $u$, we can determine all the relevant mutual information 
as 
\begin{equation}
	\left\{I(\tilde b^{\rm i}_{u,\kappa})
	\right\}_{\kappa=1}^{m_0+m_u},
	\left\{I(\tilde b^{\rm q}_{u,l})
	\right\}_{l =1}^{n_0+n_u}.
	\label{eq:Is}
\end{equation}

\subsection{Rate Region of PxQAMA}
\label{sec: rate_analysis}
We now determine the achievable rate region of the proposed system.  
The following parameters are adjustable to provide different rates for 
different users. 
\begin{itemize}
    \item \textbf{Bit-to-user Assignment}: 
	    The bits from $\calS_0$ can be assigned to different users:
\begin{equation} 
	\calS_0 = \calS_{0,1} \cup \calS_{0,2},
\end{equation} 
where $\calS_{0,u}$ denotes the collection 
of bits assigned to user $u$ in $\calS_{0}$. Let $\calA_u$ denote the collection of all bits assigned to user $u$. Under the PxQAMA framework, we have 
\begin{equation} 
	\calA_1 = \calS_{0,1} \cup \calS_{1}, \quad
	\calA_2 = \calS_{0,2} \cup \calS_{2}.
\end{equation} 
Each receiver $u$ can identify its own bits from $\tilde s_u$
based on $\calA_u$ and the relationship in \eqref{rece_bi} and \eqref{rece_bq}.
Based on the mutual information in \eqref{eq:Is}, 
the achievable rate of user $u$ is 
\begin{equation}
R_u = \underset
{\tilde b^{\rm i}_{u,\kappa}\in\calA_u} {{\sum}}
	I(\tilde b^{\rm i}_{u,\kappa}) +
\underset{\tilde b^{\rm q}_{u,l}\in\calA_u}{\sum} 
	I(\tilde b^{\rm q}_{u,l}).
    \label{eq: userrate}
\end{equation}

    \item \textbf{Constellation Optimization}: As shown in \cite{ZhZZ2023QAMA}, optimized H-QAM constellation is better than uniform QAM constellation.
	    Hence the proposed transmitter needs to optimize the 
		distances in $\calD_0, \calD_1, \calD_2$ 
		to expand the rate region.

    \item \textbf{Beam Direction Optimization and Power Allocation}: 
    Optimization of beamforming vectors $\{\bbp_0,\bbp_1,\bbp_2\}$ can be decomposed into the optimization of each beam's power and direction.
    The beam direction of $\bbp_0$ can be adjusted
		by varying $\theta_0$ in the range of $[0,\Theta]$. 
    The power of each beam is adjusted by varying
		$\{\alpha_0,\alpha_1,\alpha_2\}$ under the constraint
		$\alpha_0^2+\alpha_1^2+\alpha_2^2=1$.
\end{itemize}

To summarize, the bit-to-user assignment, the distance parameters of the constellations, and 
the precoding vectors all affect the user rates.
The proposed system first dynamically selects suitable modes 
for constellation sizes and bit assignments according to user deployments.
For the given constellation size and bit assignment, equation \eqref{eq:re_y_simp} shows that 
the beam direction-related parameter $\theta_0$ and power-related parameters $\alpha_0,\alpha_1,\alpha_2$ jointly control the equivalent channel gains of shared and private symbols for two users.
By jointly varying $\theta_0,\alpha_0,\alpha_1,\alpha_2$ and 
distance parameters in $\calD_0,\calD_1,\calD_2$, 
H-QAM constellations with different received distance parameters
$\{\tilde{d}^{\rm i}_{u,\kappa},\tilde{d}^{\rm q}_{u,l}\}$ are observed by each
user. Hence, various rate points $(R_1, R_2)$ are achieved by jointly adjusting
the parameters mentioned above. 
Finally, the achievable rate region is determined as the \textit{convex hull} of all
rate points from individual parameter configurations. 

\section{Numerical Results}
\label{sec:result}

In this section, we compare the achievable rate region of different access 
schemes in the two-user case with some specific channel realizations.
We set $N_{\rm t}=2$ and consider channel vectors as 
\begin{equation}
\bbh_1^H =\lambda_1[1, 0], \quad
	    \bbh_2^H =\lambda_2[\rho, \sqrt{1-|\rho|^2}],
    \label{eq:userchannels}
\end{equation}
where $\lambda_1,\lambda_2$ and $\rho$ control the channel strengths and the
channel correlation between two users, respectively. 
As a result, the reference SNR of user $k$ is defined as
\begin{equation}
    \gamma_u = \frac{\lambda_u^2}{\sigma^2}.  
\end{equation}
All schemes in comparisons are summarized as follows.
\begin{itemize}
    \item DPC with Gaussian inputs (DPC-Gaussian): Based on the algorithm in
	    \cite{HV_2003JASC_DPCsimu}, we determine the capacity region
		achieved by DPC with Gaussian inputs.
    \item PxQAMA: Following steps described in section \ref{sec:
	    rate_analysis}, we determine the achieve rate region of the
		proposed scheme. Corresponding to the simple receiver processing, the 
            data rates of bit-interleaved coded modulation (BICM) with H-QAM are evaluated.
            To simplify the search, the following constraints are imposed in PxQAMA systems: 
		\begin{itemize} 
			\item Each branch of $s_0,s_1,s_2$ adopts a uniform PAM constellation, i.e., $d^{\rm i}_{i,\kappa}=2d^{\rm i}_{i,\kappa+1}, d^{\rm q}_{i,l}=2d^{\rm q}_{i,l+1}$. 
	\item $s_1$ and $s_2$ are drawn from constellations with the same size, i.e., $m_1=m_2,n_1=n_2$, although the corresponding distance parameters may differ. 
		\end{itemize} 

    \item 1-Layer RSMA: As a promising multiple access technique, the 1-Layer RSMA design with both SIC-based and non-SIC receivers are compared, denoted as RSMA-SIC and RSMA-NSIC, respectively. Then the constellation-constrained (CC) rates of RSMA systems with uniform QAM are evaluated based on the derivation described in \cite{ZSB_24_NSIC}, where nonbinary coded modulation (CM), achieving performance gain with the cost of receiver complexity, is involved.
    Note that zero-forcing (ZF) precoding is adopted for private streams, while the direction of the precoder for the common stream and the power allocated to each stream are exhaustively searched to determine the region.
    \item SDMA: As a special case of the proposed scheme with $m_{\rm 0}=n_0=0$,
	    SDMA is considered as a benchmark scheme. The data rates of BICM
		are evaluated.
    \item NOMA: By entirely encoding the message for the user with worse
	    channel conditions into the common stream and turning off its
		private stream in RSMA systems, the rate region of NOMA is
		obtained. SIC-based receivers and nonbinary coded modulation are
		assumed as in RSMA-SIC.
\end{itemize}

Among all schemes in comparison, PxQAMA offers the lowest receiver complexity, 
benefiting from the elimination of SIC and the independent bit-wise demodulation process.
In the simulations, the received constellation for all the users
is no larger than H-64QAM. 
To this end, the combinations of different constellation sizes for each symbol are considered in all schemes.
Then the rate region is determined as the convex hull of all rate points from individual combinations by the Matlab function $\texttt{convhull()}$. 
\begin{figure*}[t]
    \centering
    \subfloat[{\footnotesize $|\rho|=0.2$}]{\includegraphics[scale=0.66]{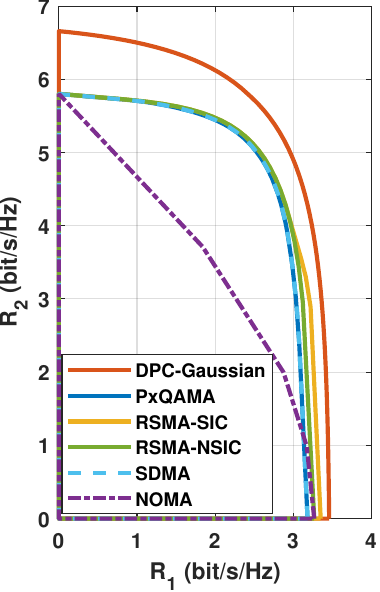}
    \label{Res1_subfig_1}}
    \subfloat[{\footnotesize $|\rho|=0.4$}]{\includegraphics[scale=0.66]{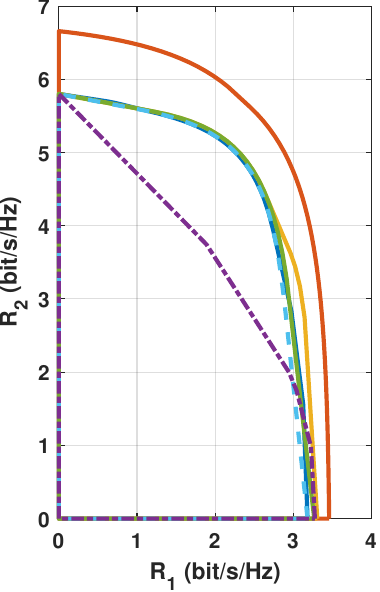}
    \label{Res1_subfig_2}}
    \subfloat[{\footnotesize $|\rho|=0.6$}]{\includegraphics[scale=0.66]{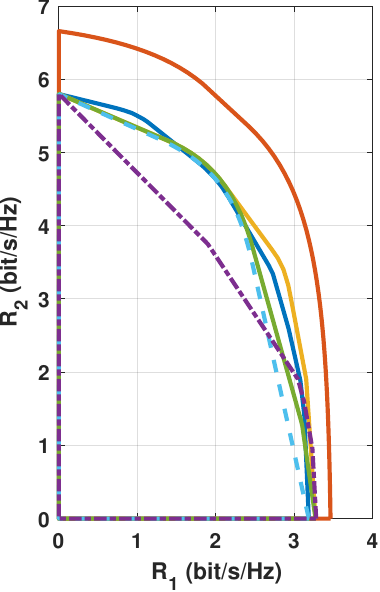}
    \label{Res1_subfig_3}}
    \subfloat[{\footnotesize $|\rho|=0.8$}]{\includegraphics[scale=0.66]{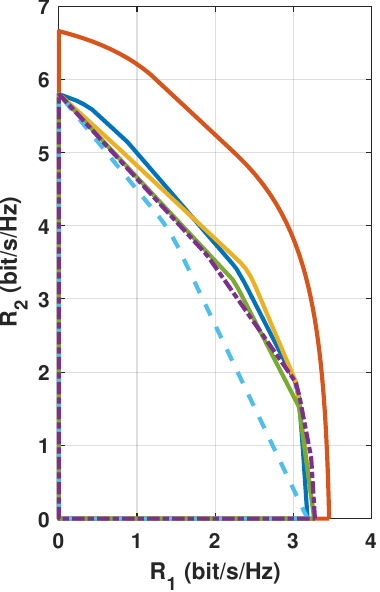}
    \label{Res1_subfig_4}}
\caption{Comparison of the rate region with ($\gamma_1, \gamma_2)=(10, 20)$ dB.}
\label{fig:Res1_fig}
\end{figure*}
\begin{figure}[t]
    \centering
    \subfloat[{\footnotesize ($\gamma_1, \gamma_2)=(0, 10)$ dB, $|\rho|=0.6$}]{\includegraphics[scale=0.62]{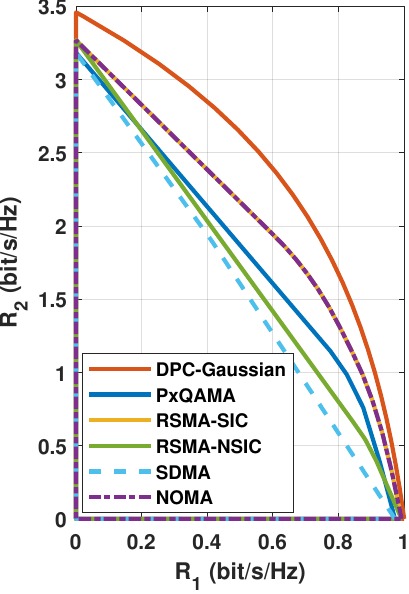}
    \label{Res2_subfig_1}}
    \subfloat[{\footnotesize ($\gamma_1, \gamma_2)=(0, 20)$ dB, $|\rho|=0.6$}]{\includegraphics[scale=0.62]{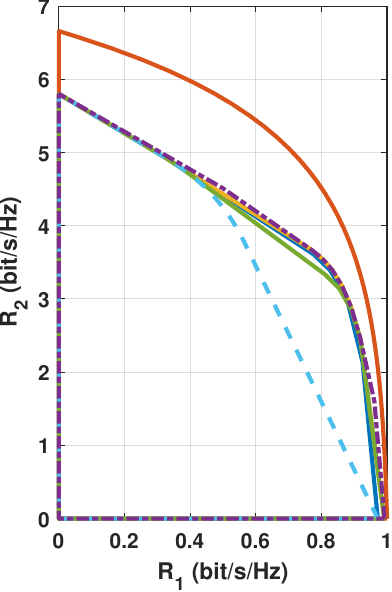}
    \label{Res2_subfig_2}}
\caption{Rate region comparison of different schemes with lower reference SNRs or larger channel strength difference.}
\label{fig:Res2_fig}
\end{figure}

\subsection{Comparison of Rate Regions}
Fig. \ref{fig:Res1_fig} compares the achievable rate regions of the proposed
system with all benchmark schemes, where reference SNRs at two users are set as
($\gamma_1, \gamma_2)=(10, 20)$ dB and various channel correlation are
considered.
Key observations are summarized as follows:
\begin{itemize}
    \item Fig. \ref{fig:Res1_fig}(a) and Fig. \ref{fig:Res1_fig}(b) show that the rate region of all schemes except NOMA almost completely overlaps when user channels are sufficiently orthogonal (low $|\rho|$), as all schemes can reduce to SDMA, which performs well in these cases\footnote{Hence the cases with relatively small $|\rho|$ are omitted in the following simulation results, and only the comparisons when $|\rho|$ is moderately high are presented to save space.}.
    Note that there are some rate gaps between corner points of different schemes, as CM rates are evaluated for RSMA and NOMA systems and BICM rates are evaluated for PxQAMA and SDMA systems.
    \item As shown in Fig. \ref{fig:Res1_fig}(c) and Fig. \ref{fig:Res1_fig}(d), the performance of SDMA degrades when the channel correlation increases.
    In contrast, NOMA achieves a good performance when user channels are sufficiently aligned (large $|\rho|$). Moreover, both PxQAMA and RSMA perform consistently well in various scenarios by dynamically adjusting the system configuration. 
    \item We further compare the rate region of the PxQAMA and RSMA systems. When $|\rho|$ is relatively large, the performance of RSMA systems degrades due to the use of non-SIC receivers instead of SIC receivers. However, PxQAMA outperforms RSMA-NSIC despite adopting a simpler receiver. Moreover, PxQAMA achieves 97.8\% and 100\% of the rate region area of RSMA-SIC when $|\rho|=0.6$ and $|\rho|=0.8$, respectively, demonstrating its effectiveness in achieving near-optimal performance even without SIC.
    Note that the rate region of PxQAMA extends beyond that of RSMA-SIC in certain areas. This is because, in PxQAMA, the rate associated with the shared symbols is not subject to the worse channel condition of the two users. By comparison, in RSMA-SIC, the common stream's rate is constrained by such conditions.
\end{itemize}

Then we further investigate the impacts of relatively lower reference SNRs and larger channel strength differences on the rate region.
Fig. \ref{fig:Res2_fig}(a)
presents a comparison of rate regions of different schemes with ($\gamma_1, \gamma_2)=(0, 10)$ dB and $|\rho|=0.6$. 
Compared with Fig. \ref{fig:Res1_fig}(c), when reference SNR decreases, the methods without SIC experience a greater rate loss than NOMA and RSMA-SIC do, despite the same channel strength differences.
It reveals that, for low SNRs, decoding interference by SIC receivers becomes more efficient than suppressing interference in the spatial domain. 
Meanwhile, among the methods with non-SIC receivers, PxQAMA still demonstrates a superior performance over RSMA-NSIC and SDMA due to the well-designed structure of the received composite constellation with Gray-mapping.
Specifically, for $R_1=0.8$ bps/Hz, PxQAMA demonstrates performance gains of $82\%$ and $33\%$ for user $2$ relative to SDMA and RSMA-NSIC, respectively.
Fig. \ref{fig:Res2_fig}(b) considers a larger channel strength difference with ($\gamma_1, \gamma_2)=(0, 20)$ dB and $|\rho|=0.6$.
Benefiting from the growth of channel strength differences, PxQAMA, along with RSMA-SIC, RSMA-NSIC and NOMA, achieves a larger rate region than SDMA in this case. 
Specifically, when $R_1=0.8$ bps/Hz, all other methods achieve a significant gain of around $119\%$ for user $2$ over SDMA.
Meanwhile, both PxQAMA and RSMA-NSIC achieve similar rate regions as the schemes with SIC-based receivers because interference and signal are more distinguishable due to significant power differences in such cases.

\begin{figure*}[t]
    \centering
    \subfloat[{\footnotesize $(12, 12)$ dB, $|\rho|=0.6$}]{\includegraphics[scale=0.65]{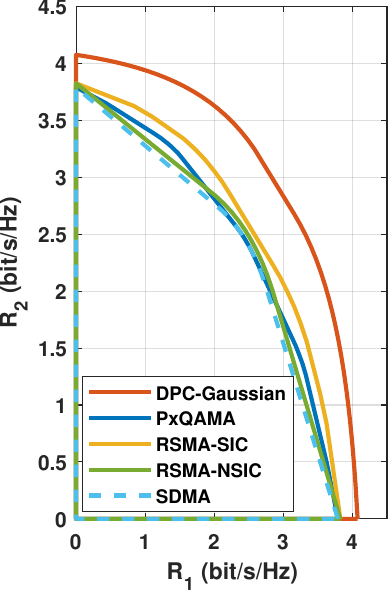}
    \label{Res3_subfig_1}}
    \subfloat[{\footnotesize $(12, 12)$ dB, $|\rho|=0.8$}]{\includegraphics[scale=0.65]{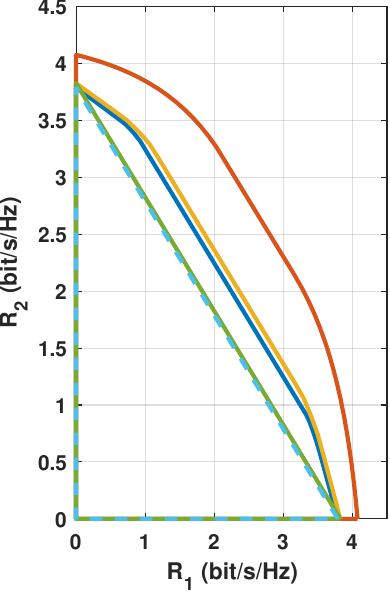}
    \label{Res3_subfig_2}}
    \subfloat[{\footnotesize $(18, 18)$ dB, $|\rho|=0.6$}]{\includegraphics[scale=0.65]{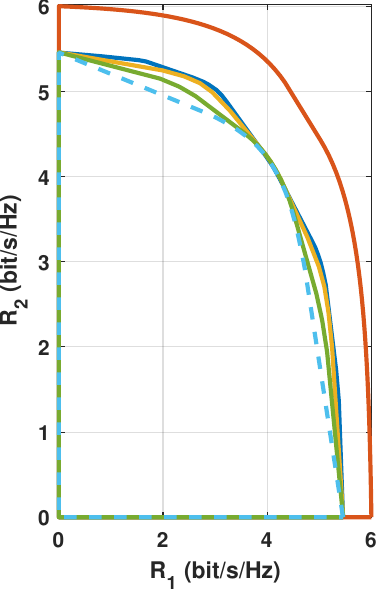}
    \label{Res3_subfig_3}}
    \subfloat[{\footnotesize $(18, 18)$ dB, $|\rho|=0.8$}]{\includegraphics[scale=0.65]{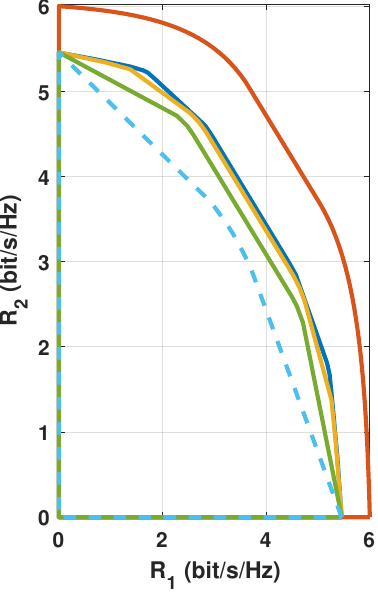}
    \label{Res3_subfig_4}}
	\caption{Rate region comparison of different schemes with equal channel strengths ($\gamma_1=\gamma_2$).}
\label{fig:Res3_fig}
\end{figure*}

We further compare the rate regions of different schemes under equal channel strength conditions, i.e., $\gamma_1 = \gamma_2$. The rate region of NOMA is omitted as it performs inefficiently when there is no channel strength disparity.
As shown in Fig. \ref{fig:Res3_fig}, both PxQAMA and RSMA-SIC outperform SDMA due to the flexible system framework.
Meanwhile, performance loss of RSMA caused by the absence of SIC receivers is observed in all sub-figures.
Compared to RSMA-NSIC, the proposed PxQAMA system achieves a larger rate region with a simpler receiver. 
Even compared to RSMA systems with SIC receivers, PxQAMA achieves almost the same rate region as shown in Fig. \ref{fig:Res3_fig}(b)--(d).

\begin{figure}
    \centering
    \includegraphics[width=0.45\textwidth]{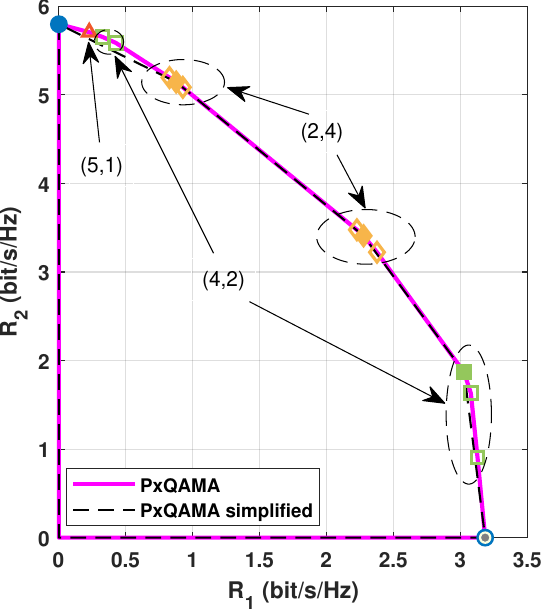}
    \caption{The two-user rate region with $(\gamma_1, \gamma_2) = (10, 20)$ dB and $|\rho| = 0.8$ is formed by 14 discrete rate points in a few separated clusters, where each point corresponds to a specific configuration for $(m_0+n_0, m_1+n_1)$. 
    A polygon formed by five selected transmission modes, each labeled by a filled marker, approximates the rate region.}
    \label{fig:trans_modes}
\end{figure}

\subsection{Selection of Transmission Modes}

The achievable rate region of PxQAMA is determined by the convex hull of many discrete rate points. 
Each rate point corresponds to 
the achieved $(R_1,R_2)$ corresponding to one transmitter configuration of the H-QAM constellations, the bit-to-user assignment, and the power splitting among the shared and private symbols. 
Each configuration is termed as a \textit{transmission mode}.

For the setting in Fig. \ref{fig:trans_modes},
a total of $N_1=14$ rate points are marked on 
the Pareto-frontier of the rate curve. 
In practice, one can reduce the number of 
transmission modes following a procedure outlined in 
\cite{ZhZZ2023QAMA}. 
Fig. \ref{fig:trans_modes} provides a demonstration with the original rate region approximated by a polygon formed by $N_2=5$ selected transmission modes: two modes of OMA, two modes with $(m_0+n_0, m_1+n_1) = (2,4)$, and one mode with $(m_0+n_0, m_1+n_1) = (4,2)$. 
Note that $m_2,n_2$ are omitted due to the assumption $m_1=m_2, n_1=n_2$.
The loss of the area formed by the polygon with five transmission modes relative to the original rate region is negligible. An upper-layer
scheduler can utilize these transmission modes to achieve a rate pair on any point of the polygon via time sharing, while always maintaining low complexity
at the receivers.

\section{Conclusion}
\label{sec:conclusion}

In this paper, we presented a downlink multiple access scheme based on joint symbol mapping of H-QAM constellations and phase-aligned precoding. 
Benefiting from the ingenious designs at the transmitter, a unique parallax effect appears in the proposed PxQAMA system, 
where each receiver observes an individual Gray-coded H-QAM constellation shaped by the user channel.
Large rate region is achieved by PxQAMA while the receiver complexity remains as low as that in an OMA system.
Hence PxQAMA shows great potential for future 6G
and IoT systems, where both receiver complexity and achievable rate are
critical metrics. 
The extension of the proposed scheme to the multi-user broadband system and the optimization with imperfect channel state information are interesting
directions for future research.

\appendix

\begin{figure*}[t]
    \centering
    \includegraphics[scale=0.75]{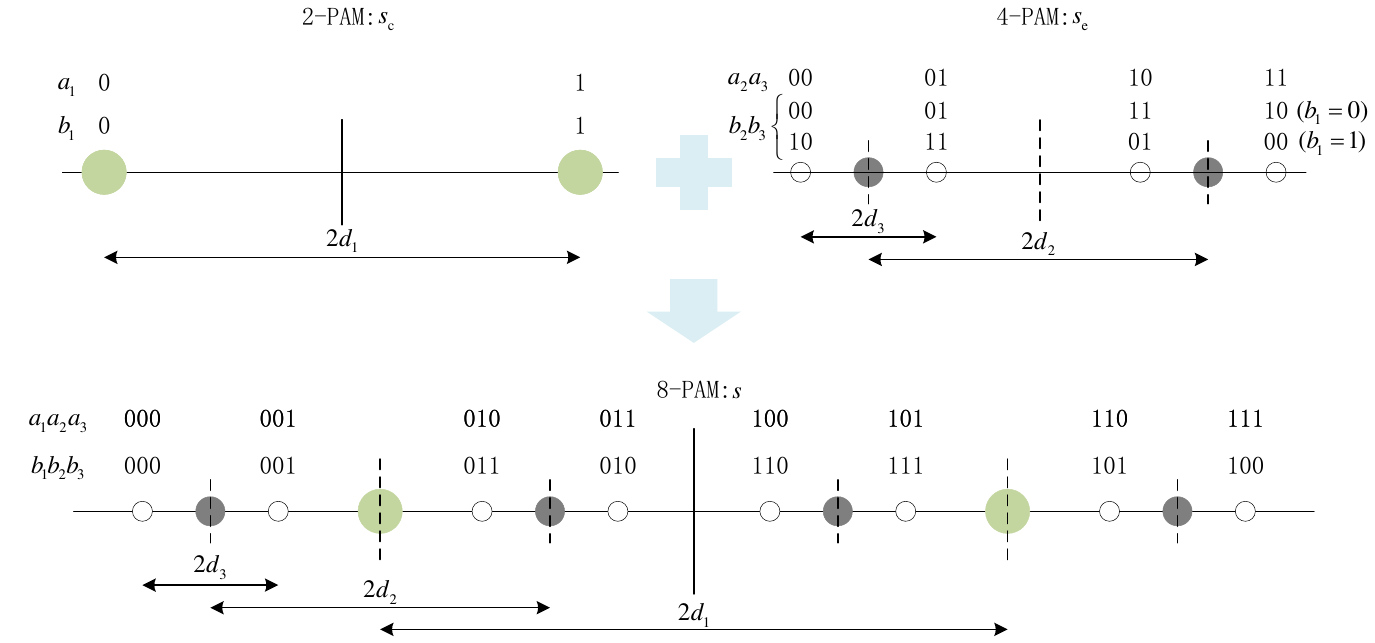}
    \caption{The hierarchical 2-PAM and 4-PAM constellation could linearly add up to a composite Gray-coded 8-PAM by joint modulation, where the mapping rule for the high-layer constellation is dependent on the bits in the low-layer constellation.}
    \label{fig:demo_insight}
\end{figure*}

\section*{Hierarchical QAM with Gray Mapping}
\label{sec:insight}

Consider a real hierarchical $M$-ary PAM constellation with distance parameters
$\{d_1, \ldots, d_m\}$, where $m=\log_2 M$.  There are a total of $M$
constellation points located at $\{\pm d_1\pm d_2 \cdots \pm d_m \}$ on the
real line, satisfying $d_k\ge 2d_{k+1}$ for $k=1,...,m-1$. Index these points with $0$ to $M-1$ from left to right. 
For a constellation point, denote the bits of 
natural mapping and the Gray mapping as
\begin{equation}
	\begin{cases}
		\text{Natural mapping:} & [a_1, \ldots, a_m]  \\
		\text{Gray mapping:} & [b_1, \ldots, b_m] 
	\end{cases}
\end{equation}
The natural mapping $\{a_1, \ldots, a_m\}$ is obtained from the binary representation of the index of the constellation point.
Eq.~(9) from \cite{1299057BER4GQAM} specifies that 
the constellation point can be computed from the bits from natural 
mapping as 
\begin{equation}
    s = \sum_{\kappa=1}^{m}(2a_\kappa-1)d_\kappa. 
    \label{gray_symbol1}
\end{equation}
Equation (1) from \cite{1299057BER4GQAM} specifies how to 
obtain the bits of Gray mapping 
$b_m$ from the bits of natural mapping via 
\begin{equation}
	\begin{split}
	b_1 &= a_1, \\
	b_{\kappa} &= a_{\kappa} \oplus a_{\kappa-1}, \kappa=2,\ldots, m,
	\end{split} 
	\label{eq:gray}
\end{equation}
where $\oplus$ represents modulo-2 addition.

Combining \eqref{gray_symbol1} and \eqref{eq:gray}, one can 
identify the constellation point via the bits of Gray mapping directly. 
First, we revert the Gray encoding in \eqref{eq:gray} as
\begin{equation}
	a_{\kappa} = b_{\kappa} \oplus b_{\kappa-1} \cdots \oplus b_{1}.
    \label{gray_convert}
\end{equation}
Second, we rewrite \eqref{gray_symbol1} as
\begin{equation}
	s = \sum_{\kappa=1}^{m} (-1)^{1+a_{\kappa}} d_{\kappa}.
    \label{gray_symbol}
\end{equation}
Combining \eqref{gray_convert} and \eqref{gray_symbol}, we obtain:
\begin{equation}
    \begin{aligned}
    s &= \sum_{\kappa=1}^{m}(-1)^{(1+\sum_{j=1}^{\kappa}b_{j})}d_{\kappa}.
    \end{aligned}
    \label{gray_decomps}
\end{equation}

The key insight based on the expression in 
\eqref{gray_decomps} is that the symbol $s$ 
can be decomposed to two symbols as 
\begin{equation}
    \begin{aligned}
	    s        &= \underbrace{\sum_{\kappa=1}^{m_0}
	    (-1)^{(1+\sum_{{j}=1}^{\kappa}b_{{j}})}d_\kappa}_{s_{\rm c}}
	    +\\
	    &\underbrace{
		    (-1)^{(1+\sum_{{j}=1}^{m_0}b_{{j}})}
		    \sum_{\kappa=m_0+1}^{m}(-1)^{(\sum_{{j}=m_0+1}^{\kappa}b_{{j}})}d_\kappa}_{s_{\rm e}}.
    \end{aligned}
    \label{gray_decomps2}
\end{equation}
With the terminology from hierarchical constellation, one 
can view $s_{\rm c}$ as a symbol at the core layer and 
$s_{\rm e}$ as a symbol at the extended layer \cite{ZL_2016_LDM}.

Fig.~\ref{fig:demo_insight} illustrates the linear superposition of 2-PAM and
4-PAM component constellations, forming a Gray-coded hierarchical 8-PAM
composite constellation.  The key is that the mapping rule for 
$s_{\rm e}$ on the extended layer is
dependent on the value of bits for $s_{\rm c}$ on the core layer. 

For H-QAM constellation, Gray coding is applied separately on the PAM
constellations on the I and Q branches. 

\bibliographystyle{IEEEtran} 
\bibliography{reference}
\end{document}